\documentclass{cernrep}
\usepackage[center]{caption}
\begin{document}
\title{Characterization of the effect of plastic gas pipes\\ with the use of a Single Wire Proportional Chamber}
\author{R. Guida, B. Mandelli, M. Corbetta, T. Marii}
\institute{CERN, EP-DT-FS, Gas Team}

\begin{abstract}
Plastic gas pipes are widely used in experimental setups for gaseous detectors, due to their flexibility and easiness of use. This report describes the characterization studies realized on the use of plastic gas pipes, particularly in relation to the accumulation of impurities in gas systems that they may cause. A laboratory setup has been implemented to evaluate the possible effects caused by plastic pipes, in terms of their specific impact on O$_2$ and H$_2$O accumulation and the consequences on detectors performance.  As Single Wire Proportional Chambers are very sensitive to the presence of pollutants and gas mixture variations, this type of detector was chosen to perform the study.
\end{abstract}


\maketitle
\vspace{5mm}
\section{Introduction}
The following work has the aim of characterizing the performance of plastic gas pipes, widely used in experimental setups for gaseous detectors. In particular, the pipe used in this study is the Festo Polyurethane Hydrolysis Resistant (FES558279, [1]), as it is the one commonly used at CERN. The main purpose is to investigate whether its use can contribute to the accumulation of O$_2$ and H$_2$O in the gas line, as they are pollutants that can have effect on detector's performance. 
\section{Experimental Setup}
\begin{figure}[h!]
\begin{center} \hspace{-15mm}
\includegraphics[width=15cm]{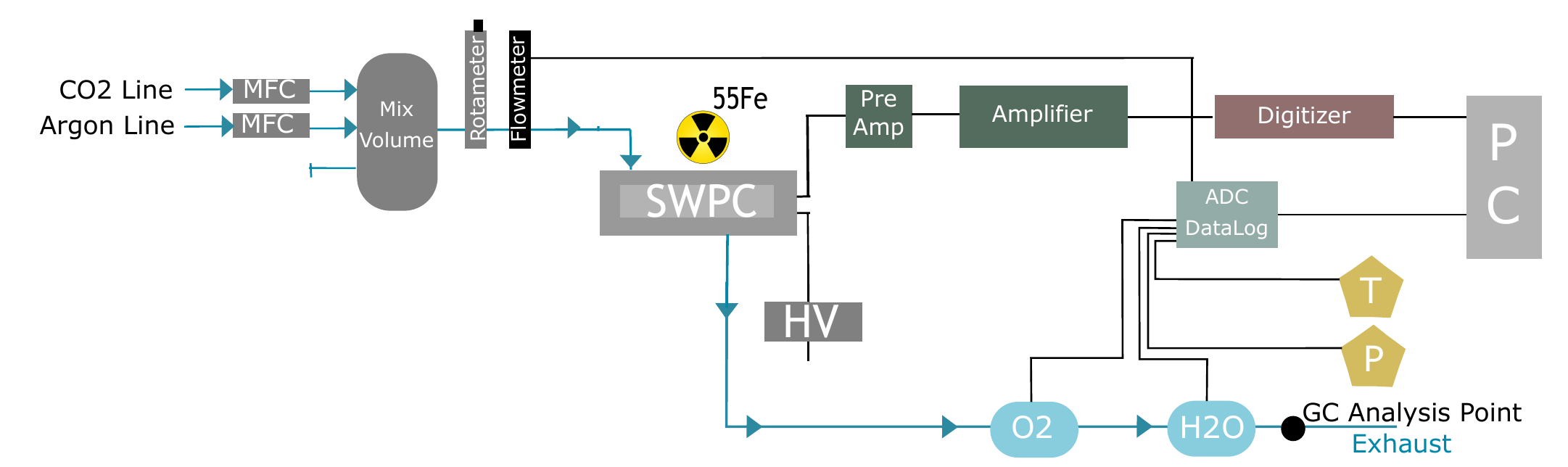}
\caption{Schematic representation of the experimental setup.}
\label{setup}
\end{center}
\end{figure}

A scheme of the experimental setup is reported in Figure \ref{setup}. A two-components mixer is used to prepare the gas mixture used for the test, Ar/CO$_2$ in proportion 70/30. After the mixer volume, a rotameter and a flowmeter are installed. The rotameter allows the gas flow regulation (from 0 l/h to 5 l/h), while with the flowmeter a continuos measurement of the flow is obtained. In this way the correct flow value can be monitored along the test, reading and recording the flowmeter output signal with an ADC Data Logger (PicoLog ADC-24 [3]). 
\\\\
The plastic pipe used in the setup is the 6 mm diameter one, and is connected after the rotameter, with Festo Quick Star Standard Push-In Fittings, in way to be able to modify its length during the test. 
\\\\
The values of O$_2$ and H$_2$O concentrations are measured at the exhaust of the gas line. The O$_2$ sensor is a O2X1 Panametrics Oxygen Transmitter ([4]), based on a galvanic fuel cell technology. It can measure O$_2$ concentration from 10 ppm up to 25\%, with user-programmable ranges. The H$_2$O sensor is a Vaisala Dewpoint Transmitter (DMT242, [5]), based on the Vaisala DRYCAP\textregistered \ thin film polymer sensor, that covers dew point measurement range of $\pm$ 60 degrees. They are connected to the ADC Data Logger, to continuously record the measured concentrations. Moreover, environmental parameters such as temperature and atmospheric pressure are recorded (Electrotherm with Pt100 sensors). 
\\\\
As the study context is the use of plastic pipes in setups for gaseous detectors, a Single  Wire  Proportional  Chamber (SWPC) is also installed along the gas line (volume 0.085 l), between the variable-length plastic pipe and the sensors. This type of detector was chosen as its performance is very sensitive to outgassing and gas mixture variations, specifically to the presence of impurities even at the level of ppm. From this also comes the choice of the gas mixture Ar/CO$_2$ (70/30), as it is the one usually used for wire chambers [2]. The SWPC is irradiated with a $^{55}$Fe radioactive source (activity of 30 MBq). The detector signal is shaped and amplified by a pre-amplifier (CAEN A1422) and amplifier (ORTEC 474 Timing Filter Amplifier) chain, and then digitized using a multichannel analyzer (CAEN Waveform Desktop Digitizer, DT57242). The pulse shape is automatically recorded every hour with a ROOT based software, to continuously monitor detector performance. 
\begin{figure}[h!]
\begin{center}
\includegraphics[width=6cm]{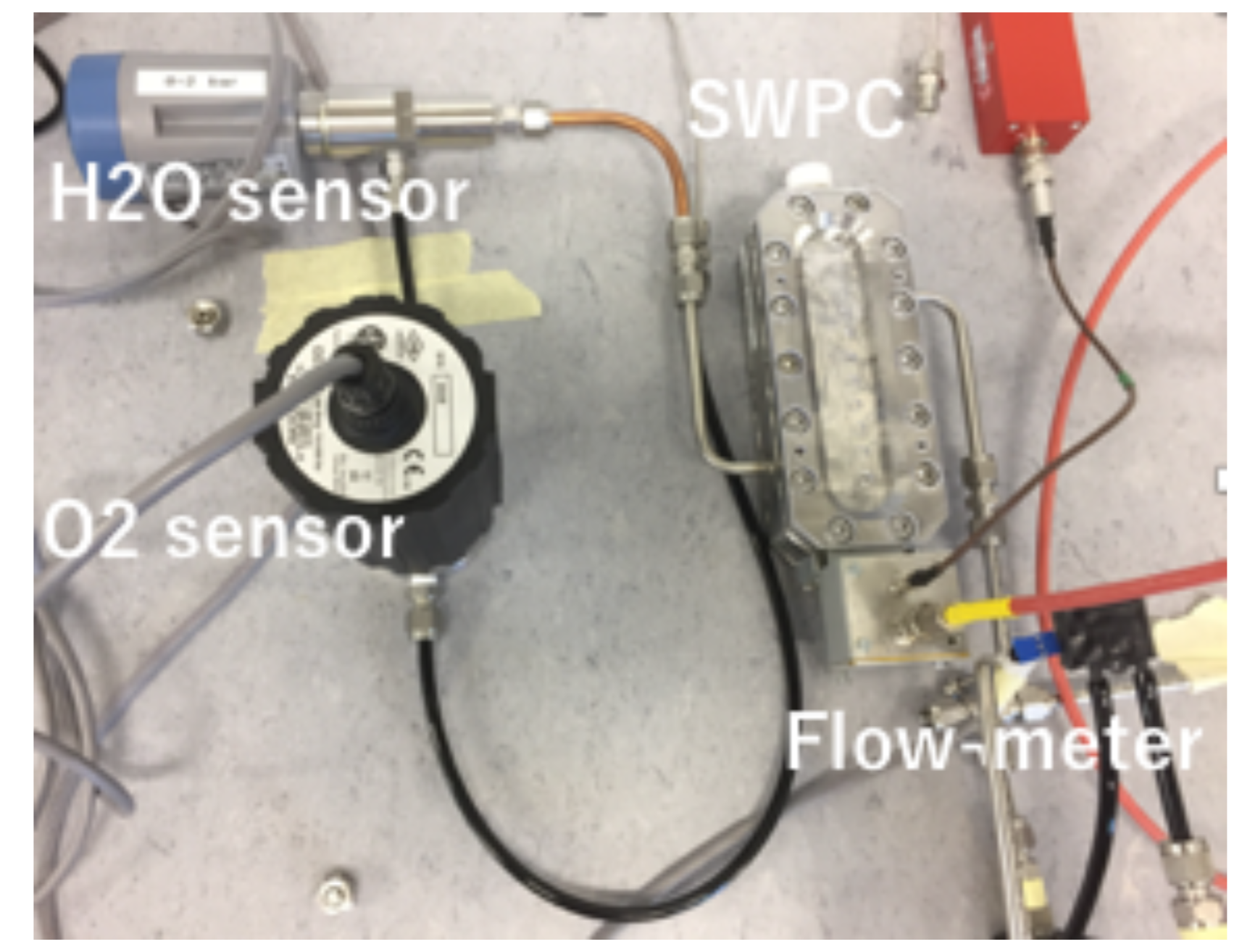} \hspace{5mm}
\includegraphics[width=3.2cm]{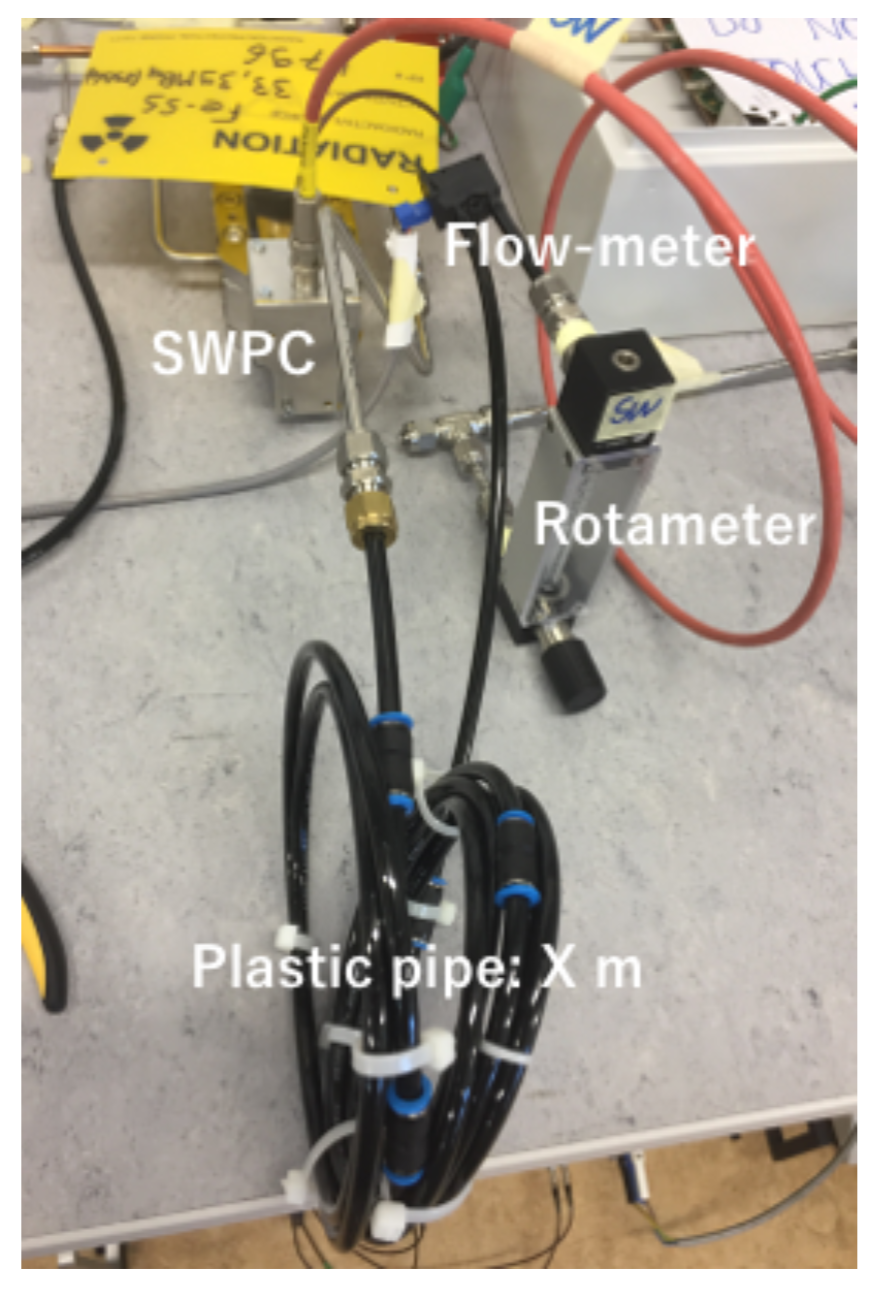}
\centering
\caption{Experimental setup, with impurities sensors and SWPC (left), \\ and plastic gas pipe with rotameter and flowmeter (right).}
\label{tubo}
\end{center}
\end{figure}

\section{Performance test of plastic pipe} 
\begin{figure}[h!]
\begin{center}
\includegraphics[width=7.3cm]{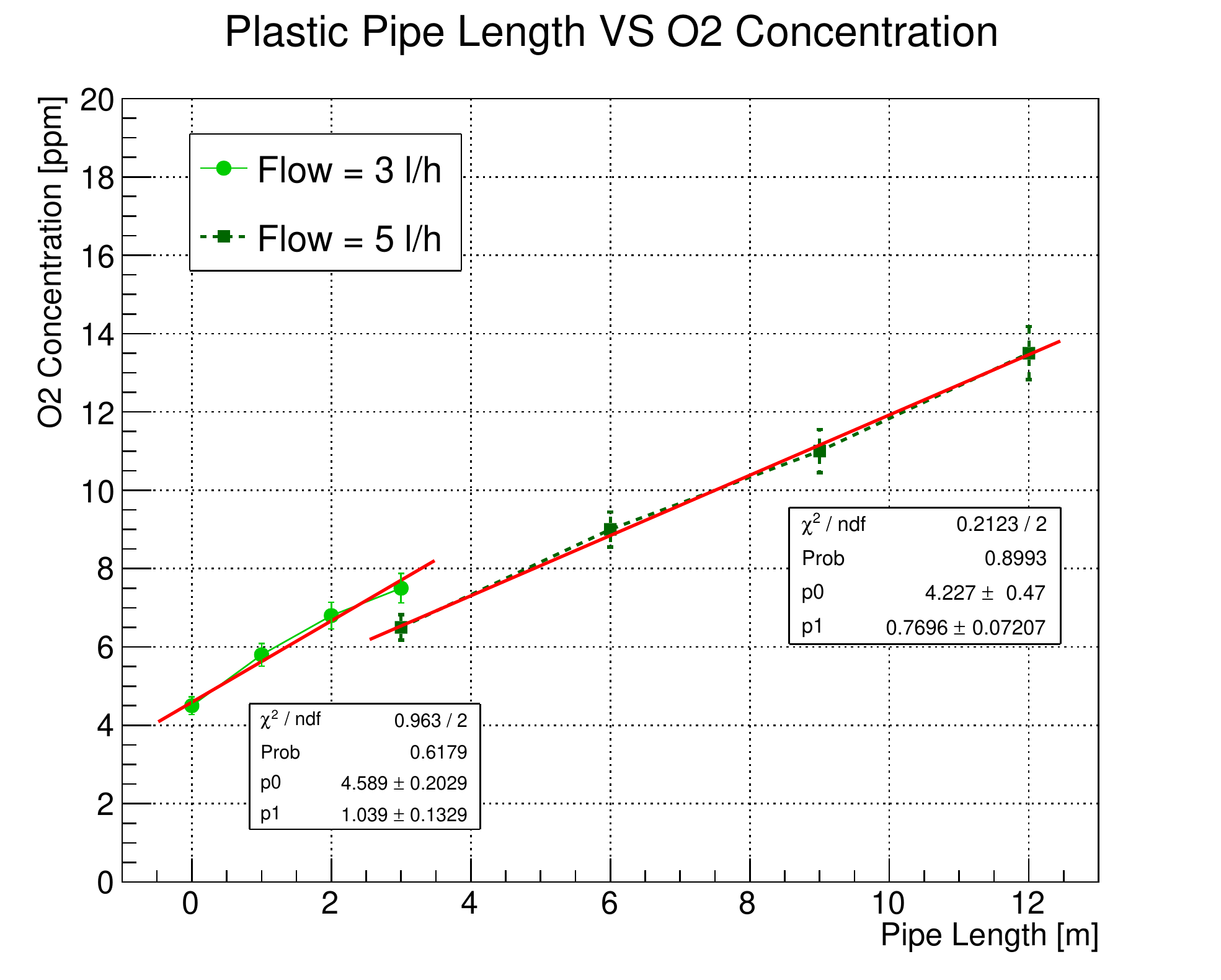}
\includegraphics[width=7cm]{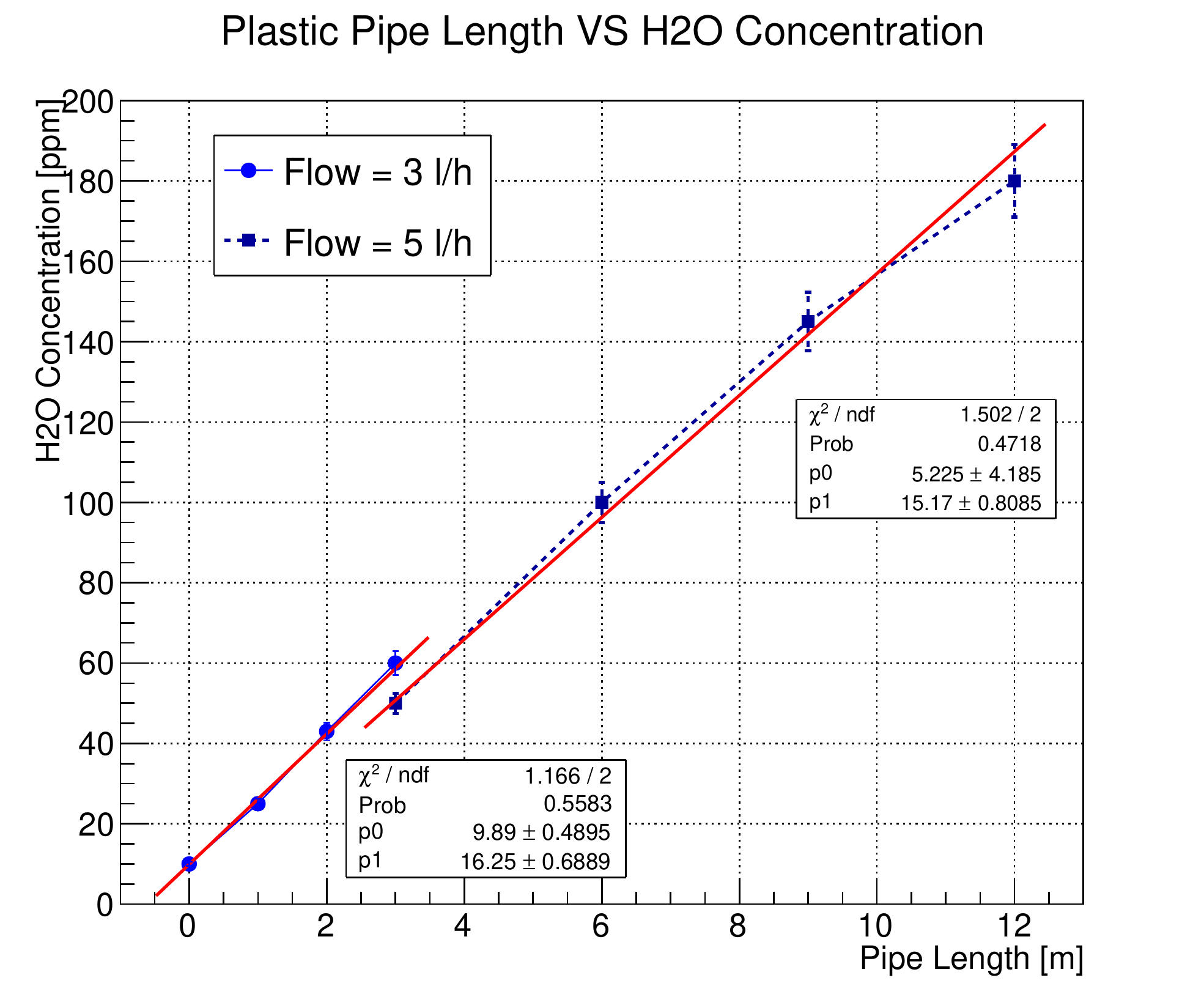}
\caption{Impurities concentration trend in relation to plastic gas pipe length.}
\label{h2o2}
\end{center}
\end{figure}

\begin{figure}[h!]
\begin{center}
\includegraphics[width=8cm]{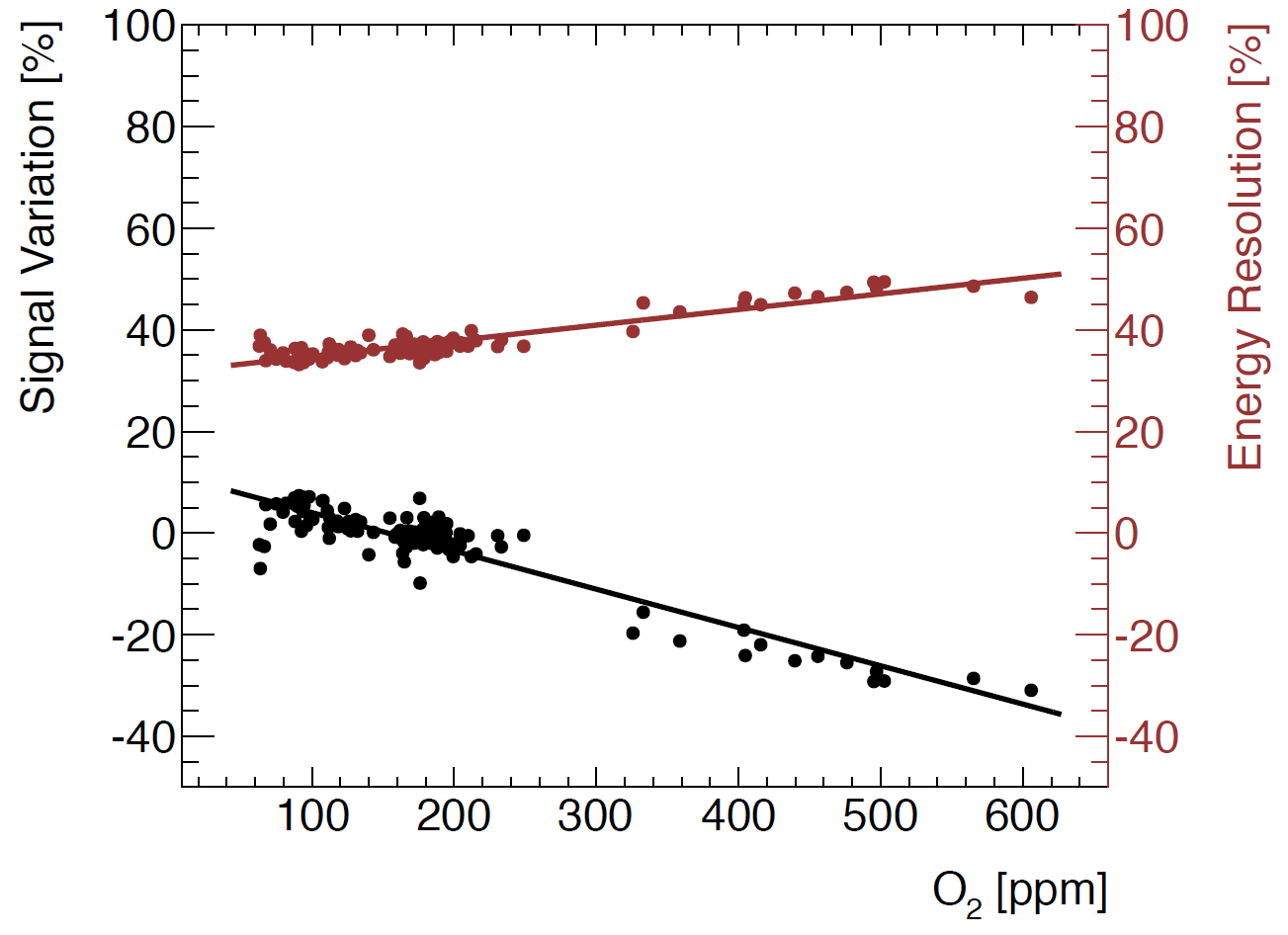}
\caption{SWPC gain for O2 concentration.}
\label{swpco2}
\end{center}
\end{figure}

\begin{figure}[h!]
\begin{center}
\includegraphics[width=8cm]{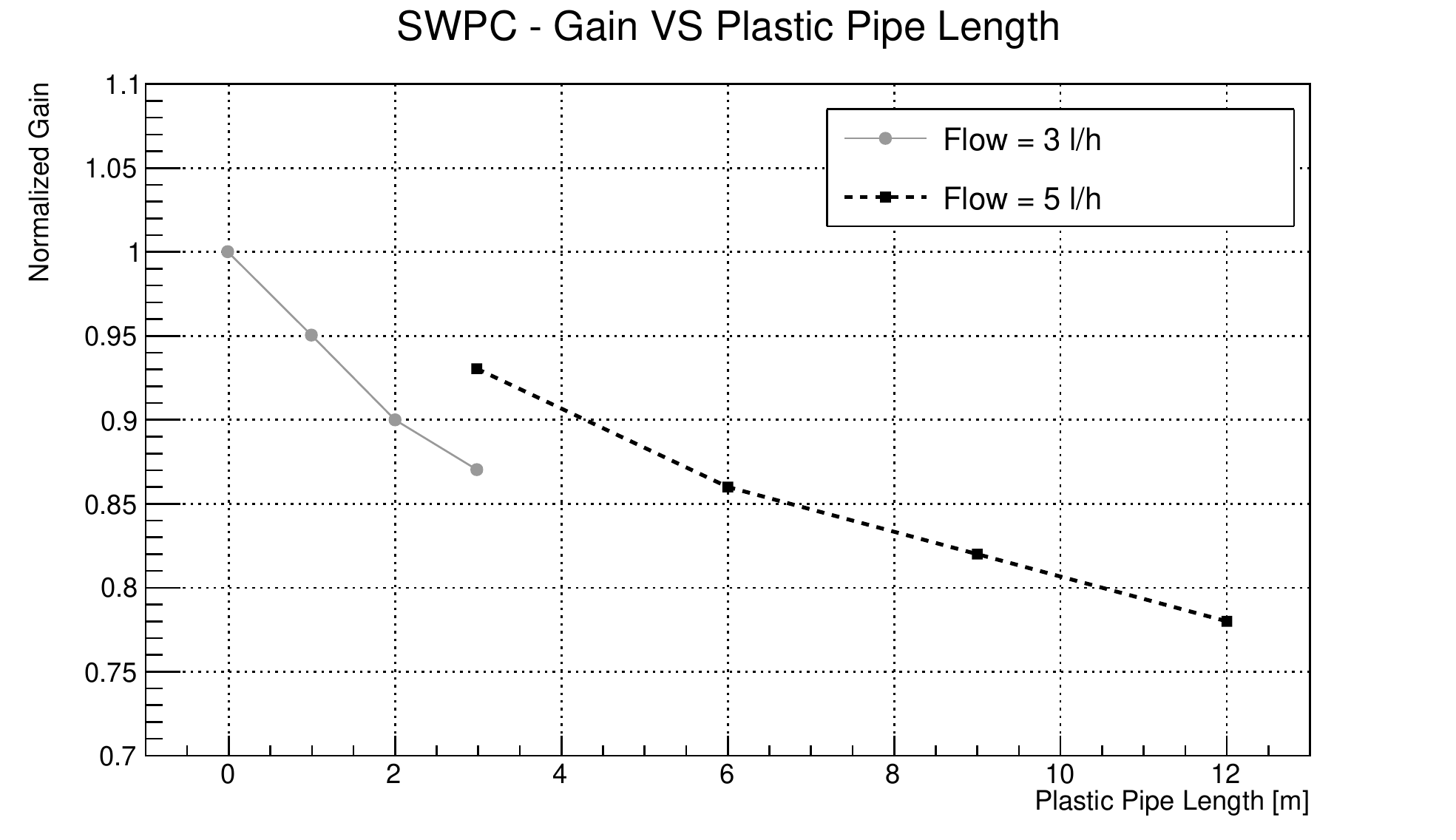}
\caption{SWPC gain for different lengths of the plastic gas pipe.}
\label{gain}
\end{center}
\end{figure}

To test the impact of plastic pipe in the system its length was progressively increased, for two different values of the input gas flow, to probe any possible change in the effect of plastic for different flow rates. In particular, they were added 1 m, 2 m, 3 m of gas pipes for an input flow of 3.0 l/h, and 3 m, 6 m, 9 m and 12 m for 5.0 l/h. The test was conducted at atmospheric pressure (970 bar) and fixed temperature (21 \textdegree C). 
\\\\
Looking at the relation between the length of plastic pipe added to the system and the concentration of O$_2$ and H$_2$O (Figure \ref{h2o2}), it can clearly be seen how they increase for greater pipe length. The O$_2$ content in the line varied from 5 ppm to 15 ppm, while the H$_2$O concentration from 10 ppm up to about 200 ppm. The measurement points taken with the same pipe length but different gas flow shows the effect of the flow increase. For both impurities, with a higher flow (from 3 l/h to 5l/h) the pollutant concentration shows a decrease of about 15\%. Moreover, looking at the linear fit of the concentration trends, in both cases the slope is lower for higher flows. 
\\\\
The amplification gain of the SWPC was also recorded for the different points. From previous tests [6] it is known that its performance is influenced by the concentration of O$_2$ and H$_2$O (an example is showed in Figure \ref{swpco2}). Given the measured variations of the two pollutants, their effect is expected to be seen in the gain of the chamber placed after the gas pipe. 
\\
\\
Figure \ref{gain} shows the result obtained in the two sets of measurements, where the gain was offline normalized for the initial value (3 l/h, without plastic pipe) and corrected for temperature and pressure. It can be seen for both data-sets that the gain is decreasing with the increase of plastic pipe length, with a loss up to the 20\% with respect to the initial value. The loss in performance is linked to the increase of the pollutants, as the decrease of the amplification gain comes fomr the combined effect of the presence of O$_2$ and H$_2$O. 
\section{Conclusions}
The aim of the reported test is to evidence possible effects of the use of plastic gas pipe in gas systems for gaseous detector studies. It has been then quantified their impact in relation to the accumulation of O$_2$ and H$_2$O, that could affect detectors performance. 
\\\\
The following table shows the values extracted from the fit of the trend of impurities concentration in relation to the length of plastic gas pipe added to the system. 
\begin{table}[h]
    \centering
    \begin{tabular}{c|c|c}
                 &  3 l/h & 5 l/h\\ \hline
       O$_2$  & 1 ppm/m & 0.77 ppm/m \\
       H$_2$O & 16.25 ppm/m & 15.17 ppm/m
    \end{tabular}
    \caption{Slop of the linear fit on the trend of impurities in relation to pipe length.}
    \label{tab:my_label}
\end{table}
\\
A first conclusion that can be drawn is that the increase in H$_2$O concentration is much more significant than the one of O$_2$. Moreover, it can be seen how the increase trend is different for the two gas flow tested. The plastic pipe seems to have a slightly lower impact if the gas flow is higher.
\\\\
Taking into account the different data-sets, it can be concluded that the addition of gas plastic pipes leads to an increase in the O$_2$ concentration of about 1 ppm per meter, and of about 15 ppm per meter for the H$_2$O concentration.

\end{document}